\documentclass[12pt]{article}
\usepackage{amsmath,amssymb,amsthm}
\usepackage{natbib,graphicx}

\newcommand{\cd}{\mbox{$\,|\,$}}

\newcommand{\p}{\mathbb{P}}
\newcommand{\E}{\mathbb{E}}

\title{Defining and estimating stochastic rate change in a dynamic general insurance portfolio}
\author{Roland Ramsahai}

\begin{document}

\maketitle

\begin{abstract}
Rate change calculations in the literature involve deterministic methods that measure the change in premium for a given policy. The definition of rate change as a statistical parameter is proposed to address the stochastic nature of the premium charged for a policy. It promotes the idea that rate change is a property of an asymptotic population to be estimated, not just a property to measure or monitor in the sample of observed policies that are written. Various models and techniques are given for estimating this stochastic rate change and quantifying the uncertainty in the estimates. The use of matched sampling is emphasized for rate change estimation, as it adjusts for changes in policy characteristics by directly searching for similar policies across policy years. This avoids any of the assumptions and recipes that are required to re-rate policies in years where they were not written, as is common with deterministic methods. Such procedures can be subjective or implausible if the structure of rating algorithms change or there are complex and heterogeneous exposure bases and coverages. The methods discussed are applied to a motor premium database. The application includes the use of a genetic algorithm with parallel computations to automatically optimize the matched sampling.
\end{abstract}

\section{Introduction}
The task of measuring rate change has traditionally been addressed by applying various deterministic methods \citep{M:2001,L:2018} to the portfolio of policies across multiple years. This discussion enhances the techniques available for such a task by proposing the use of statistical methods which recognize the stochastic nature of the final premium charged for a policy. For a particular policy, the uncertainty arises from actuarial and underwriting judgment \citep{V:2004}, market conditions and other sources in the complex process of arriving at the final premium exchanged. The aim here is similar to previous actuarial advances that extended classic deterministic reserving methods to stochastic reserving techniques which recognize the uncertainty in the claims generating process \citep{EV:2002}. In addition to providing more accurate estimates, the statistical techniques provide measures of uncertainty around the estimates, with a full predictive distribution for rate change. To exploit statistical approaches, the discussion provides a novel definition of rate change as a statistical parameter to be estimated from observed data and not simply as just some property to be measured in the data.

The philosophy acknowledges that there is sampling variability in the data and that the true rate change is not observed in a finite sample. Therefore estimators are developed which converge to the true rate change. The formulation of rate change in a statistical framework unlocks a large body of literature on parameter estimation with measures of estimator uncertainty and statistical tests. The methods are essentially required to estimate premium trends after adjusting for the changes in the characteristics of the policies being written. This is shown to be achievable with well known regression methods but such methods require assumptions about the relationship between the premium charged, policy year, exposure and other policy terms and conditions. This is not always feasible so the focus here is on the use of matched sampling methods \citep{S:2010}, which do not require these assumptions.

Matched sampling adjusts for changes in policy characteristics by meticulously constructing a sample of similar policies in different years. The resulting nonrandom sample then reflects premium trends for a policy with constant characteristics. The approach is relatively unknown in the actuarial literature but has been thoroughly extended and applied in areas such as Epidemiology \citep{BSRGAS:2006}, Econometrics \citep{HIT:1998} and others \citep{HIKS:2007}, to name a few examples. Matching is commonly used to adjust for the differences in demographics across groups to allow a fair comparison of treatments or schemes. It is necessary because the demographic characteristics are related to the treatment or action assigned and confounds their relationship to outcomes of interest \citep{GRP:1999}. In this discussion, their use is proposed to adjust for the differences in the mix of policy characteristics across portfolios when estimating rate change. The mix of policy characteristics vary with policy year and influences the premium charged, so it confounds their relationship. The approach comes fully equipped with algorithms for efficient automated implementation, statistical checks of results and a large literature with practical advice and technical details \citep{S:2010}.

To address portfolio changes from cancellations, alterations or new business, the actuarial literature has mainly focused on re-rating policies or re-stating premiums as if written in a different year. These approaches rely on various recipes for adjusting the policy premium charged to a different year than when it is actually written \citep{M:2001,B:2009}, and can be subjective to some extent. The empirically based matching methods avoids such conjectures by finding similar policies that were written at different times. The similarity of policies is quantified by multivariate measures of the deviations in deductibles, limits, exposures and other terms and conditions. The matching algorithms then produce a sample of policies, in which there is the same mix of characteristics across years or the policy characteristics are balanced across years. In practice, the policy mix is not perfectly balanced across years but there are computationally intensive algorithms that optimize balance and checks for whether the balance achieved is satisfactory.

Section~\ref{sec:matchdemo} describes a very simple example to demonstrate the use of matched sampling in the estimation of rate change. The rate change parameter is defined using standard probabilistic notation in Section~\ref{sec:probdef} for individual policies and aggregate portfolios. An intuitive justification of the matched sampling estimator is given in Section~\ref{sec:intjust} and its practical implementation is discussed in Section~\ref{sec:practimpl}. Rate change estimation with regression is described in Section~\ref{sec:regest}. An application of the various techniques is given in Section~\ref{sec:motapp}, with R code \citep{RCT:2017} for implementation in Appendix~\ref{sec:softimp}. The discussion throughout is primarily on estimating rate change between two years but Section~\ref{sec:discussion} briefly comments on extending the ideas to multiple years.

\section{Simple demonstration of matching concept}\label{sec:matchdemo}

\subsection{Original dynamic portfolio}
Consider the portfolio of policies in Table~\ref{tab:matchsimp} in which the deductible is the only defining policy characteristic. The list of policies in the portfolio in a particular year, $T=0$, and the subsequent year, $T=1$, are shown in the table. The portfolio is dynamic since the mix of deductibles changes between the two years. The deductible for policy d is increased at the end of the earlier year, policies (f,j) are not renewed and policies (k,l) are new policies which are written in the later year.
\begin{table}\centering
\begin{tabular}{|c|c|cccccccccccc|}\hline
\multicolumn{2}{|c|}{Policy}&a&b&c&d&e&f&g&h&i&j&k&l\\\hline
Deductible&Earlier ($T=0$)&1&10&10&2&5&2&10&2&1&2& &  \\\cline{2-14}
(\$mil)		&Later ($T=1$)  &1&10&10&5&5& &10&2&1& &2&10\\\hline
\multicolumn{14}{c}{}\\\cline{1-2}
\multicolumn{2}{|c|}{Matched pairs}&\multicolumn{12}{c}{}\\\hline
Policy&Earlier ($T=0$)&a&b&c&e&e& &g&h&i& &j&g \\\cline{2-14}
			&Later ($T=1$)  &a&b&c&d&e& &g&h&i& &k&l\\\hline
\end{tabular}
\caption{Example showing a dynamic portfolio (top) with changes to the deductible and the matched sample (bottom).}
\label{tab:matchsimp}
\end{table}
It is necessary to use the portfolios to estimate the historical rate change for the portfolio at time $T=1$, which is the premium trend if the later portfolio was written in both years. A simple comparison of the premium charged for policies a, b, c, f, g, h and i gives an indication of rate change for these policies since they are renewed without changes. However this does not reflect the rate change for the entire portfolio in the later year.

\subsection{Matched sampling}
Matched sampling estimates historical rate change for the portfolio in the later year by matching policies with no direct counterpart in the earlier year to one with the most similar deductible. Policy k at $T=1$ is matched to policy j at $T=0$ and policy l at $T=1$ is matched to policy g at $T=0$ since they have the same deductibles. Interestingly, policy d at $T=1$ is matched to policy e, not d, at $T=0$ since its deductible was changed. The renewed deductible of $\$5$mil on policy d makes it more similar to the old policy e than the old policy d with a deductible of $\$2$mil. All pairs of matched policies, including those which were paired with their old unaltered versions, are combined to create the matched sample. The resulting matched sample consists of all policies in the later year and their matched policies in the earlier year. Unmatched policies in the earlier year are ignored since the aim is to estimate rate change for the later portfolio.

For simplicity, policy k was just matched to one policy j but it could have also been matched to d or f. There were also multiple best matches or ties for policy l as well. In more complex cases, the process for dealing with ties and the number of matches must be specified. The old policy e was matched to multiple policies but it is sometimes necessary to require matching without replacement in the pool of potential matches. This example is also particularly simple since a zero difference in the deductible is the only criteria for a match. More general cases involve matching on multivariate metrics on multiple variables with some non-zero threshold for defining an acceptable match.

\subsection{Analysis of matched sample}
Ideally, the matched sample should be a static portfolio in which the mix of policy characteristics is the same in both years. In this example, the matched sample includes the original $T=1$ portfolio and a comparison $T=0$ portfolio with the same mix of deductibles. The similarity of deductible distribution between years is called the balance and has clearly improved in the matched sample relative to the original portfolio. In this simple case, perfect balance is achieved but only approximate balance is usually achieved in practice. It therefore becomes necessary to use further diagnostic checks such as quantile-quantile plots, weighted differences across policies and other statistics that quantify the contrast in the distribution of policy characteristics across years.

The steps in matched sampling become more complex when multiple variables are involved but the matching process essentially aims to balance the multiple characteristics. Once the balance is deemed to be acceptable, rate change is estimated in the matched sample. The estimator is appropriately risk adjusted because the achieved balance implies that the policy characteristics are identical across years.

\section{Probabilistic notation and definitions}\label{sec:probdef}

\subsection{Definition of premium charged as a random variable}
The written policy premium charged is defined as a random variable $W$ to reflect uncertainty in the process of arriving at its final value. Policies with the same terms may achieve different premiums in the same year because of the market and economic climate. The policy characteristics is defined as a random variable $X$ to reflect the distribution of policy types in a portfolio. The variable $X$ could be univariate or multivariate to represent a combination of characteristics such as limit and exposure. The quantity $\p(X)$ is the probability of a policy having specific characteristics when randomly chosen from all policies written in all years. The distribution $\p(X\cd T)$ is then the mix of policy types for the portfolio of policies written in year $T$. The quantity $\E(W\cd X,T)$ is the average premium actually charged for a policy written with the terms $X$ in year $T$. For simplicity, $T$ is defined as binary for most of the discussion, where $T=1$ if the policy is written in the later year and $T=0$ for the earlier year.

\subsection{Individual rate change}
The change in expected premium for a particular policy across years is defined as the individual rate change, since it applies to an individual policy as defined by the specific terms and conditions $X$. Let the individual rate change be defined as
\begin{align}
	\delta_{X}=g(\mu_{X1},\mu_{X0}),	\label{eq:dxdef} 
\end{align}
where $\mu_{XT}=\E(W\cd X,T)$, $g(\cdot)$ is some function that measures contrast, such as a difference $\mu_{X1}-\mu_{X0}$ or a relativity $\mu_{X1}/\mu_{X0}$. The quantity $\delta_X$ captures the change in expected premium for a policy with terms $X$ when written in the later year $T=1$ relative to the earlier year $T=0$.

For policy $X$, it is the individual prospective rate change for year $T=0$, but the individual historical rate change for year $T=1$. The policies being compared have the same terms and thus reflects the same risk so $\delta_X$ is risk adjusted. The quantity $\delta_X$ can be used to onlevel the premium charged for policy $X$. For example, the predicted onlevelled premium at $T=1$ for policy $X$ written at $T=0$ is $\hat{\mu}_{X1}=\hat{\mu}_{X0}\hat{\delta}_{X}$ if $\delta_X$ is a ratio, where $\hat{\mu}_{X0}$ is the premium charged at time $T=0$. The onlevelled premium is a predicted value since there is uncertainty in the premium charged at time $T=1$.

\subsection{Aggregate rate change}
The change in expected premium for the aggregated portfolio of all policies is the aggregate or average rate change, since it applies to the group of policies as defined by the mix of terms and conditions in all years. In order to define aggregate rate change, it is necessary to define the notation $\mu_{T|T'}=\E_{X|T'}(\mu_{XT})$ and $\mu_{T}=\E_{T'}(\mu_{T|T'})=\E_X(\mu_{XT})$. The parameter $\mu_{T|T'}$ is the average premium charged if the portfolio written in year $T'$ had been written in year $T$. This is because it is the expected premium charged in year $T$ for the mix of characteristics in the portfolio in year $T'$. Similarly the parameter $\mu_{T}$ is the average premium charged if all policies had been written in year $T$. In the causal inference literature, $\mu_{T}$ is a causal quantity that represents the average premium given an intervention in which $T$ unnaturally arises independently of $X$. This is contrasted with the usual quantity $\mu_{T|T}=\E(W\cd T)$, which is the average premium conditional on an observation of $T$ that is naturally dependent on $X$ \citep{D:2002}. The latter is the premium charged in year $T$ since it is trivially the average premium charged if the portfolio written in year $T$ had been written in year $T$.

The aggregate rate change parameter for the aggregated portfolio across all years is defined as $\delta=g(\mu_{1},\mu_{0})$. However insurers are usually interested in the average rate change for the portfolio in a particular year and not the aggregate of all years. The aggregate rate change for the portfolio written in year $T$ is defined as
\begin{align}
	\delta^{T}=g(\mu_{1|T},\mu_{0|T})	\label{eq:dtdef}
\end{align}
since $\mu_{T'|T}$ reflects the pricing in year $T'$ for the portfolio in year $T$. For example, $\delta^{1}$ is the average historical rate change for the later portfolio at $T=1$. Also, the predicted onlevelled premium at $T=1$ for the portfolio written at $T=0$ is $\hat{\mu}_{1|0}=\hat{\mu}_{0|0}\hat{\delta}^0$ if $\delta^T$ is a ratio, where $\hat{\mu}_{0|0}$ is the average premium charged in year $T=0$. By iterated expectations, $\delta=\E_T(\delta^T)$, which agrees with intuition since the average rate change across all years is the average of the rate change for the portfolios in each year.

\section{Methods and practical considerations}\label{sec:practimpl}
This section describes the use of matched sampling to estimate the rate change parameter of interest. An intuitive explanation of the validity of the procedures described here are given in Appendix~\ref{sec:intjust} but proper proofs are quite lengthy and can be obtained in the literature \citep{AI:2006,AI:2016}.

\subsection{Selecting matches}
Matched sampling to estimate rate change involves the selection of the entire portfolio of interest, together with matching policies in a comparison portfolio. If $\delta^1$ is of interest then the entire portfolio in year $T=1$ should be selected. Policies should then be selected from the portfolio of policies written in year $T=0$ to match each policy in year $T=1$ . Policies in year $T=0$ that are not chosen as matches are irrelevant and not included in the matched sample. Policies in year $T=1$ that cannot even be approximately matched are also dropped but are more consequential. If too many are dropped then the portfolio mix in the matched sample would not resemble the original mix in year $T=1$ and the estimator would be biased for $\delta^1$. This is because $\delta^1$ is the aggregate rate change for the portfolio mix in year $T=1$. More generally, estimating $\delta^t$ requires the selection of all policies in year $T=t$ and their matches. Estimating $\delta$ requires searching for matches for each policy in both years $T=1$ and $T=0$.

It is not necessary to select matching policies based on all terms and conditions in policies. The only relevant characteristics are those that are considered to be both related to the premium charged and the year written. It is important to balance these confounding characteristics across the years since they distort the premium trends across policy years. In practice, characteristics that are only weakly related to the premium charged could be ignored. This is because, even if they are strongly related to the year written, they remove little bias between years but greatly decrease the statistical efficiency \citep{BSRGAS:2006}.

In ideal cases, selecting matches based on the similarity of multiple policy characteristics can be equivalently replaced with selection on a single characteristic called the propensity score \citep{RR:1983}. If the portfolio in year $T=1$ is of interest then this score is defined as $\p(T=1\cd X)$, the probability of a policy with characteristics $X$ being written in year $T=1$. It provides a suitable quantitative summary measure of the characteristics of a policy. In practice, the true propensity score is unknown so the propensity score is usually estimated by a logistic regression, but the matching estimator is still asymptotically valid \citep{HIT:1998}. Propensity score matching selects matches for policies that have a similar chance of being written in a particular year, regardless of whether their individual characteristics are similar, and is easier to implement since it involves a single characteristic. However, since there is sampling variability and approximate matching in practice, the literature recommends matching on the propensity score as well as the relevant individual policy characteristics \citep{RR:1985}.

Most applications involve selecting one best match but multiple matches could be selected in cases where there may be multiple acceptable policy matches that are not necessarily the best. Similarly, all policies that tie for the best match could be included or ties could be randomly broken to decide which policy to select as a match for the sample. Both considerations involve an assessment of whether the increase in bias is worth the improvements in efficiency by sampling more policy matches. Selecting matches with replacement may be necessary if there are only a few policies to find matches among but it risks the chance of over representation of some policies in the matched sample. These and other practicalities are discussed in detail in \citet{S:2010} and its references.

\subsection{Quantifying the similarities between policies}
In practice, there may be no policies in the comparison portfolio which have exactly the same characteristics as the policy in portfolio to be matched. Thus it is necessary to find approixmate matches and define some quantitative measure of the similarity between two policies. This allows the selection of best matches by assessing tradeoffs in similarities in individual policy characteristics. This can be done by implementing matching to minimize the Mahalanobis distance (MD) \citep{R:1980}. This is a scalar quantity that measures the multivariate distance between individuals in two groups. It allows the matching of policies across portfolio years in order to minimize the MD metric, which would minimize the overall difference in the mix of policy types. The MD between two policies i and j with a vector of policy characteristics $X$ is
\[	\textsc{md}(X_i,X_j)=\sqrt{(X_i-X_j)^TS^{-1}(X_i-X_j)},		\]
where $S$ is the sample covariance matrix of $X$ and $X^T$ is the transpose of the matrix $X$. The MD metric is a function of quantitative characteristics so ordered qualitative characteristics could be given a numeric label to reflect their ordering and then characteristics could be matched with the nearest label. Otherwise, qualitative characteristics that are important to match on, whether they have an ordering or not, should be matched exactly if policies with different values are not comparable. This exact matching would make the contribution of the characteristic to MD be zero. The propensity score variable is included alongside all other characteristics in the MD metric.

The MD is the most common metric used for matching in practice. However there is no guarantee that minimizing the MD achieves the best attainable balance of portfolio mix across policy years. Some other metric may have been more optimal. \citet{S:2011} has suggested the more flexible generalization of MD for achieving optimal balance
\begin{align}
	\textsc{gmd}(X_i,X_j,W)=\sqrt{(X_i-X_j)^T\left(S^{-1/2}\right)^TWS^{-1/2}(X_i-X_j)}, \label{eq:gmd}
\end{align}
where $W$ is a $k\times k$ positive definite matrix and $S^{-1/2}$ is the Cholesky decomposition of S, i.e.\ $(S^{-1/2})^TS^{-1/2}$. The GMD metric gives different weights to the deviations in each of the individual policy characteristics, after standardizing by the sample variances. However MD gives the same weight to all. It is then possible to choose the weight matrix $W$ that achieves the best balance of portfolio mix when minimized.

The matching process can be implemented with calipers that define certain thresholds for the maximum value of the similarity metric that is deemed an acceptable match. A caliper of zero is equivalent to exact matching.

\subsection{Algorithms for optimized matching}
Classic algorithms for the implementation of matched sampling involve iterative refinements of the matching options and the propensity score. After each iteration, if the balance is assessed to be unsatisfactory then the next iteration is adjusted to vary the variables being matched, the model used to estimate the propensity score or the other matching options that determine the number of selected matches. Although this approach involves manual refinement of the matching options, it has been successfully applied in many applications. Even single step applications without any refinement can result in appropriately matched samples. However some datasets may require multiple iterations with no guarantee that the balance improves after each iteration.

More advanced methods for iterative refinement of the matched sample involve automated computations. \citet{S:2011} describes an automated genetic algorithm to optimize balance by searching the space of all $W$ to find the GMD metric, as defined in Eq.~(\ref{eq:gmd}), which results in the optimal balance of when minimized. It guarantees asymptotic convergence to the optimal sample of balanced portfolios. The search for an optimal $W$ essentially finds the weighting that should be given to deviations in each of the policy characteristics. It also effectively chooses which variables to match on since a zero weight implies that a characteristic is not being considered in selecting matches. Any difference in the characteristic between two policies would have no contribution to the GMD metric that quantifies the similarity of two policies.

The genetic algorithm for automated balance optimization of \citet{S:2011} has been shown to improve the findings of published analyses that use classic matching approaches \citep{RGS:2011}. It is designed to propose batches of solutions for $W$, called generations, which asymptotically evolve toward a batch containing the optimal solution for $W$. The number of proposed solutions in each generation, called the population size, is fixed and increasing the population size usually improves the final solution. The design is such that each subsequent generation has better solutions than the previous one. Further technical details are left to the references.

\subsection{Assessing the matched sample}
Since policies are only matched approximately, it is necessary to assess the matched sample to determine whether it is satisfactory. This is done by checking whether the mix of various policy types is balanced or similar across years. Various metrics and plots are used to assess the similarity. These include test statistics and p-values from statistical tests of the equality of the distributions of policy characteristics in different policy years. Some test statistics that are commonly used are likelihood ratios, standardized mean differences and Kolmogorov-Smirnov (KS) test statistics \citep{A:2009}.

The KS-test statistic is the maximum deviation of the cumulative distribution between years and considers deviations across all moments of the policy type distribution. However, the other metrics only test for deviations in the mean of the distribution. The smaller the test statistic, or the larger the corresponding p-values, the better the balance. Quantile-quantile plots are also useful for comparing the policy distributions across policy years. It is also necessary to confirm the balance of the policy characteristics between the matched sample and the original portfolio of interest. Otherwise the matched sample may not be representative of the mix of policy types in the portfolio of interest.

\section{Regression based estimators}\label{sec:regest}
It is possible to use generalized linear models as an alternative to matching methods for estimating rate change. However they require assumptions about the relationship between the expected premium charged $\mu_{XT}$, the policy characteristics $X$ and the year written $T$. The inclusion of $X$ in the predictor adjusts for changes in policy characteristics. The estimated coefficients in a regression can only interpreted as a measure of rate change in certain cases. An additive linear predictor will be considered but the comments can be extended to more complex models.

\subsection{Estimating individual rate change}
Consider the simple model for the individual premium charged for a policy
\begin{align}
	h(\mu_{XT})=\beta_{\emptyset}+\beta_X+\beta_T	\label{eq:indpremglm}
\end{align}
where $X$ and $T$ are categorical, $h(\cdot)$ is the link function, $\beta_{\emptyset}$ is the intercept and $\beta_X$ and $\beta_T$ are the coefficients of $X$ and $T$. This model relates the expected premium charged across different years and can be directly used to onlevel premium if the coefficients are known. For a particular individual policy $X$, replacing the year with a different policy year than its actual written year, gives the onlevelled premium.

If $h(\cdot)$ is the identity link function and $g(\cdot)$ in Eq.~(\ref{eq:dxdef}) is the difference function then $\delta_X=\beta_1-\beta_0$ and is the same for all policies as defined by $X$. Therefore $\hat{\delta}_X=\hat{\beta}_1-\hat{\beta}_0$ is an estimator of individual rate change, where $\hat{\beta}$ is the estimator of $\beta$ in the generalized linear model. Similarly, if $h(\cdot)$ is the log link function and $g(\cdot)$ in Eq.~(\ref{eq:dxdef}) is the ratio function then $\hat{\delta}=\exp(\hat{\beta}_1-\hat{\beta}_0)$ is an estimator of individual rate change. Here again, the estimator is the same for all policies. It is only possible in special cases such as these to interpret the coefficients as measures of rate change that are constant for all policies. If the identity link was used with a ratio or the log link was used with a difference then the rate change would depend on the policy characteristics $X$.

For the additive predictor in Eq.~(\ref{eq:indpremglm}), if $h(\cdot)$ is the identity link function then the time dependent loading is a fixed amount for all policies written in a given year. In contrast, if $h(\cdot)$ is the log link function then the time dependent loading is a fixed proportion for all policies written in a given year. The choice of link function can be based on judgment about which of these or other implications are most appropriate. If the time dependent component of the linear predictor $\beta_T=\beta T$ is a linear function of policy year, then the coefficient $\beta$ can be interpreted as a simple or continuously compounded rate change under the linear or log link functions respectively. The inclusion of an interaction in the predictor would invalidate these interpretations and simplifications.

\subsection{Estimating average rate change}
The model in Eq.~(\ref{eq:indpremglm}) represents the relationship between premium charged and policy characteristics at the individual level. At the aggregate level the average premium charged for the portfolio in year $T'$ is
\[  \mu_{T|T'}=\E_{X|T'}\{h^{-1}(\beta_{\emptyset}+\beta_X+\beta_T)\}	\]
from Eq.~(\ref{eq:indpremglm}). If $h(\cdot)$ is the identity link function and $g(\cdot)$ in Eq.~(\ref{eq:dtdef}) is the difference function then the average rate change for the portfolio in year $T'$ is $\delta^{T'}=\beta_1-\beta_0$ and can be estimated by $\hat{\delta}^{T'}=\hat{\beta}_1-\hat{\beta}_0$. The value of the average rate change $\delta^{T'}$ is the same regardless of the portfolio year $T'$. This is intuitive since the individual rate change does not depend on the policy characteristics $X$, assuming $g(\cdot)$ in Eq.~(\ref{eq:dxdef}) is also a difference function. Similar reasoning also confirms the fact that the average and individual rate change are equal, $\delta^T=\delta_X$.

If $h(\cdot)$ is the log link function and $g(\cdot)$ in Eq.~(\ref{eq:dtdef}) is the ratio function then $\hat{\delta}^{T'}=\exp(\hat{\beta}_1-\hat{\beta}_0)$ is an estimator of average rate change. The value of the average rate change is again the same for all years $T'$, because the average and individual rate change are equal and do not depend on the policy characteristics.

\section{Application to private motor insurance data}\label{sec:motapp}
This section describes an application to data, with the R code for implementation in Appendix~\ref{sec:softimp}.

\subsection{Description of data and outcome}
The methods discussed are used to estimate rate change from the premium database of a private motor portfolio of an unknown French insurer. The dataset is obtained from the CASdatasets package \citep{DC:2018} in R statistical software \citep{RCT:2017}. Although the data was collected from January 2003 to June 2004, rate change is estimated from policy year 2003 ($T=0$) to 2004 ($T=1$). This analysis aims to estimate historical rate change for the portfolio in 2004. For simplicity, only a subset of 5000 policies were used in the analysis, with $2248$ in the portfolio in $2004$ to be matched to the other policies in $2003$.

The variables {driver age}, {driver gender}, {bonus malus}, {vehicle age}, {vehicle power}, {vehicle class}, {number of vehicles on the policy}, {policy region} and {type of garage} are the policy terms which are assumed to be confounders. They are related to the premium charged and also distributed differently between the portfolios in 2003 and 2004. Driver gender is included as a confounder on the basis that there was no mandate for gender neutral pricing in 2003 and 2004. The vehicle power variable was converted to a numeric variable, with labels that kept the ordering, to allow approximate matching.

\subsection{Implementation settings}
All of the matching methods were required to match exactly on the vehicle class, number of vehicles and type of garage since these were considered as very strong predictors of the premium charged. Only policies in which these terms are exactly the same could be considered as comparable. Additionally, policy region was matched exactly because it is a confounder and could not be matched approximately, since it could not be reasonably converted to a numerical variable. Exact matching excludes all policies in 2004, that do not have an identical counterpart in the 2003 portfolio, from the sample used in estimating rate change. Thus it is necessary to ensure that only a small number of policies are dropped from the sample.

Multiple matched sampling methods are applied to the dataset. The propensity score was estimated using a generalized linear model with a logistic link function and an additive linear predictor. Classic matching finds the best matches on the individual variables, including the linear predictor of the propensity score. It is common practice to use the linear predictor instead of the actual propensity score because it is a valid substitute and is on a linear scale relative to the variables being matched. Propensity score matching finds the nearest match on the linear predictor of the propensity score only. Complete matching is applied to find policies with all terms and conditions being identical. Computational matching is implemented to optimize matches on the propensity score and all individual policy characteristics. The simple bootstrap algorithm described in \citet{AS:2014} was used for estimating sampling variability for the rate change estimates from the matched sampling methods.

Computational matching here uses a standard optimization function that maximizes the smallest p-value in order to achieve the best balance of the mix of policy types. However, it is possible to define more complex functions, such as the user defined function in \citet{RGS:2011} which allows matching according to some specified order of prioritization for policy characteristics. While the computational matching involves intensive computations, the software implementation allows parallel computation to reduce computing time. The application here was implemented on a single desktop by using parallel computation with a multi-core processor on multiple threads with the R parallel package \citep{RCT:2017}. The computation time was less than an hour.

A linear model on the log scale was used as a regression alternative to matching methods for estimating rate change. This application used an additive linear predictor and assumed that the premium charged follows a lognormal distribution. The predictors were the same policy characteristics as those used for pairing in the matched sampling methods. A bootstrap approach was used to calculate the confidence interval for the rate change estimates.

\subsection{Results}
The balance of the distribution of policy types between the years 2003 and 2004 in the matched portfolio was assessed by p-values from t-tests for binary policy characteristics and KS-tests for others. By design, all of the matching methods achieve a minimum 100\% p-value on the vehicle class, number of vehicles, type of garage and policy region. This is a useful improvement when compared to a minimum p-value of 22\% in the original portfolios for these policy characteristics.

Table~\ref{tab:appresbal} shows the p-values for policy terms that were approximately matched, to assess the balance before and after matching. 
\begin{table}[ht]\centering
\begin{tabular}{|cl|ccccc|}\hline
&Policy&\multicolumn{5}{|c|}{p-values (\%)}\\\cline{3-7}
&characteristics&Before&Complete&Classic&PScore&Computational\\\hline
     &driver age&22.1  &100		  &87.7 	&58.2  &81.7 \\
  &driver gender&83.6  &100		  &1.6  	&97.3  &49.6 \\
    &bonus malus&28.1  &100		  &44.1 	&49.5  &51.1 \\
    &vehicle age&13.7  &100		  &94.6 	&88.6  &92.6 \\
  &vehicle power&11.0  &100		  &91.7 	&86.2  &97.1 \\\hline
 &matched number&2248  &144		  &2084		&2084  &2084\\\hline
\end{tabular}
\caption{Table of p-values for assessing the balance of policy characteristics between 2003 and 2004, before and after complete, classic, propensity score and computational matched sampling.}
\label{tab:appresbal}
\end{table}
Relative to the original portfolios, classic matching resulted in better balance on all characteristics except driver gender. Propensity score matching essentially improved the balance on all policy characteristics. Computational matching further improved the balance on all variables except driver gender. However it achieved the largest minimum p-value across all policy characteristics, which is the criteria being optimized by the genetic algorithm.

Although the application of complete matching resulted in a sample with exactly the same policy mix in 2003 and 2004, over 93\% of the 2248 policies in the 2004 were excluded from the estimation. Only 7\% were excluded for the approximate matching methods. This demonstrates the impracticality of methods which idealistically require all comparisons to involve identical policies. The issue is even worse when estimating rate change only for renewals as expiring, as that does not allow the pairing of policies in 2003 with those in 2004 that are not renewals. The regression approach distinguishes itself from matched sampling since it uses the entire portfolio in 2003 and 2004 for estimation of rate change. It does not drop any of the policies and therefore ignores any issues with the overlap of predictors across policy years.

Estimates and confidence intervals for aggregate rate change for the portfolio in 2004 are given in Table~\ref{tab:appresout} for each of the methods considered.
\begin{table}[ht]\centering
\begin{tabular}{|cl|cccccc|}\hline
&&\multicolumn{6}{|c|}{Rate change (estimates/confidence intervals)}\\\cline{3-8}
&                   &Before   &Exact    &Classic  &PScore   &Computational &Regression\\\hline
              &total&7.4      &7.7      &6.6      &5.9      &5.7      &5.1\\
&										&(4.5,11) &(5.3,9.6)&(4.5,8.4)&(3.6,8.5)&(3.8,7.8)&(3.5,6.5)\\
																																					&&&&&&&\\
							&legal&9.2      &7.6      &8.9      &7.9      &8.2      &7.6\\
&										&(7.1,11) &(5.2,9.9)&(7.2,10) &(6.2,9.8)&(6.6,9.8)&(6.3,8.9)\\
																																					&&&&&&&\\
&mandatory liability&8.7      &6.8      &9.4      &7.4      &7.8      &6.9\\
& 									&(5.4,12) &(4.2,9.1)&(7.3,11) &(4.7,10) &(5.6,10) &(5.5,8.2)\\\hline
\end{tabular}
\caption{Table of estimates and 95\% confidence intervals of historical rate change for the portfolio in 2004, from regression and complete, exact, classic, propensity score and computational matched sampling.}
\label{tab:appresout}
\end{table}
All of the methods differ a lot from the original portfolio, showing that adjustment for the mix of policy types was necessary. The propensity score and computational matching estimates for total premium rate change are similar and less than the exact and classic matching estimates. This shows that the very small sample used for the exact method and the balance achieved by classic matching was not sufficient. The regression estimates are different but their validity relies on the model assumptions and they may even be more reliable since they are obtained from the entire original portfolio. Similar comments explain the discrepancy in rate change for the legal and mandatory liability premium.

The confidence intervals roughly reflect an uncertainty of about 2\% above and below each of the estimates. This is useful to consider when making decisions informed by rate change or onlevelling premiums. It reflects uncertainty in the pricing environment.

\section{Discussion}\label{sec:discussion}
Matching and regression both have relative advantages, depending on whether it is more important to utilize the entire portfolio of policies or avoid assumptions about the relationship between premium and policy characteristics. It is argued here that matching is more relevant in insurance applications for calculating rate change. In certain cases, if there was already some generalized linear model available for pricing then this may contain some component that can be interpreted as some rate change quantity. However, pricing algorithms are usually more complex than a linear model. They also provide an average loss cost, not the average premium charged, which incorporates other factors in addition to loss cost.

The discussion only includes regression and matched sampling estimators. Another popular approach in the causal inference literate for the type of adjustment required is inverse probability weighting \citep{RHB:2000}. It weights each policy by the inverse of the propensity score, which essentially adjusts for the different mix of policy types across years. For example, if a particular set of policy characteristics is twice as likely to be written in a particular year then all policies of that type in that year will be given a weight of one half.

The choice of estimator is not always mutually exclusive. It is possible to apply regression techniques in the sample selected from matching. This could be useful since it further adjusts for any difference in the mix of policy types that was not eliminated by approximate matching. A weighted regression application with weights defined by the inverse of the propensity score is a hybrid of inverse probability weighting and regression. While the propensity score is by far the most common score used in matched sampling, other scores such as the prognostic score \citep{H:2008} also exist and can be additionally matched to reduce discrepancies in policy types across years.

The regression methods are straightforward to extend if more than two years are compared. It is even possible to extend the matching methods, but it involves multiple matching applications. Matches for the portfolio of interest can be found in all years and the premium trend in the matched sample would reflect rate change. Rate change across multiple years could be a comparison across the beginning and end year or some combination of rate changes for each consecutive pair. Each matching application could result in different policies being dropped in the portfolio of interest but the aim is to minimize this distortion. For the portfolio at time $n$, the quantities of interest in the period of years $1,\ldots,N$ would be $\delta^{n}_{2,1},\ldots,\delta^{n}_{N,N-1}$, where $\delta^{T}_{T'',T'}=g(\mu_{T''|T},\mu_{T'|T})$ is the rate change for the portfolio in year $T$ from year $T'$ to $T''$.

The application did not estimate rate change for various other coverages. For example, there are certain policies in which the fire premium is zero, so the ratio of premiums is infinite. The denominator premium is not zero, it is missing since that coverage was not included. Such issues may be overcome by fitting a two part model that models the probability of being missing and the premium conditional on not being missing.

\appendix

\section{Sketch reasoning for matching policies}\label{sec:intjust}
Let all policies in the original portfolio be indexed with positive integer labels $\mathbb{Z}_+$, where renewals are distinguished with separate indices, and $\mathcal{I}_{XT}=\{i\in\mathbb{Z}_+:X_i=X,T_i=T\}$ be the labels for the policies with characteristics $X$ and written in year $T$. Similarly let $\mathbb{Z}^*_+\subset\mathbb{Z}_+$ be the multisubset of labels for the policies in the matched sample obtained by selecting the policies in $\mathcal{I}_{T'}$ and their matches on $X$ from the other policies $\mathcal{I}\backslash\mathcal{I}_{T'}$, where $\mathcal{I}_{T}=\cup_T\mathcal{I}_{XT}$. The collection of labels for the policies in the matched sample with characteristics $X$ and written in year $T$ is $\mathcal{I}^*_{XT}=\{i\in\mathbb{Z}^*_+:X_i=X,T_i=T\}$.

\subsection{Complete exact matching to balance portfolios}
The original portfolio can be considered as a random sample. Therefore the averages and proportions, for policies in the original portfolio which were written in year $T$ with terms $X$, would asymptotically converge to the true means and probabilities
\begin{align}
 \begin{array}{c}	\hat{p}_{XT}\rightarrow p_{XT},\,\,\,\,\hat{\mu}_{XT}\rightarrow \mu_{XT},\,\,\,\,
							 \hat{\mu}_{T|T}\rightarrow\mu_{T|T} \end{array}	\label{eq:xtconv}
\end{align}
for all $T$, where $\hat{p}_{XT}=|\mathcal{I}_{XT}|/|\mathcal{I}_{T}|$, $\hat{\mu}_{XT}=1/|\mathcal{I}_{XT}|\sum_{i\in\mathcal{I}_{XT}}W_i$, $\hat{\mu}_{T|T}=\sum_{i\in\mathcal{I}_T}W_i$, $p_{XT}=\p(X\cd T)$ and $|\cdot|$ is the cardinality of a set. The premium charged for a particular policy in the matched sample is the same as it was in the original portfolio. Therefore, conditional on $X$ and $T$, the distribution of $W$ is the same in the matched sample as the original portfolio. This implies that $\hat{\mu}^*_{XT}\rightarrow \mu_{XT}$, from Eq.~(\ref{eq:xtconv}), where $\hat{\mu}^*_{XT}=1/|\mathcal{I}^*_{XT}|\sum_{i\in\mathcal{I}^*_{XT}}W_i$ is the empirical estimator in the matched sample.

In the ideal case, all of the policies in $\mathcal{I}_{T'}$ are matched exactly to others in $\mathcal{I}\backslash\mathcal{I}_{T'}$ with identical characteristics $X$. Therefore the resulting matched sample has the same mix of policies in all years as year $T'$ of the original portfolio. This means that $\hat{p}^*_{XT}=\hat{p}_{XT'}$ and hence $\hat{p}^*_{XT}\rightarrow p_{XT'}$ for all $T$, from Eq.~(\ref{eq:xtconv}), where $\hat{p}^*_{XT}=|\mathcal{I}^*_{XT}|/|\mathcal{I}^*_{T}|$. It follows that
\begin{align}
	\hat{\mu}^*_{T|T}\rightarrow\mu_{T|T'}	\label{eq:muttest}
\end{align}
since $\hat{\mu}^*_{XT}\rightarrow \mu_{XT}$ and $\hat{\mu}^*_{T|T}=\sum_X \hat{\mu}^*_{XT}\hat{p}^*_{XT}$, where $\hat{\mu}^*_{T|T}=\sum_{i\in\mathcal{I}^*_T}W_i$. This means that the average premium in each year $T$ in the matched sample, $\hat{\mu}^*_{T|T}$, is a valid estimator of $\mu_{T|T'}$. It can then be loosely reasoned that $g(\hat{\mu}^*_{0|0},\hat{\mu}^*_{1|1})$ is a valid estimator of $\delta^{T'}$ in Eq.~(\ref{eq:dtdef}). The estimator is simply a function of the average premium in each year of the matched sample since the matching process adjusts for any differences in the portfolio mix. This explanation focuses on the empirical estimator but it is reasoned that regression or other estimators in the original sample can be applied to the matched sample without any adjustment for $X$. It is usually straightforward to obtain an asymptotically valid estimator of $\mu_{T|T}$ or $g(\mu_{0|0},\mu_{1|1})$ in the original portfolio since $\mu_{T|T}=\E(W\cd T)$.

\subsection{Incomplete and multiple matching}
The explanation so far has relied on the assumption that exact matching is achieved for all policies in year $T'$ in the original portfolio. However this is unlikely to be achieved in portfolios in practice, particularly if $X$ represents many terms and conditions that can have a wide range of values. There are various checks for whether the matched sampling has been successful, by assessing whether the approximations $|\mathcal{I}^*_{XT}|\approx|\mathcal{I}^*_{XT'}|$ and $|\mathcal{I}^*_{XT'}|\approx|\mathcal{I}_{XT'}|$ are reasonable, for all $X$ and $T$. The former is related to the balance of characteristics across years in the matched portfolio and the latter is related to the portion of policies in year $T'$ for which appropriate matches were found. If the approximations are considered reasonable then $\hat{p}^*_{XT}\approx\hat{p}_{XT'}$ since $|\mathcal{I}^*_{T}|=|\mathcal{I}^*_{T'}|$, for all years $T$.

The estimator of $\mu_{T|T'}$ in Eq.~(\ref{eq:muttest}) is only valid if the same number of matches are included for each policy in year $T'$. However if tied matches are included then there may be a different number of matches per policy in the matched sample. The estimator $\hat{\mu}^*_{T|T}$ in Eq.~(\ref{eq:muttest}) is adjusted by replacing the definition of $\hat{p}^*_{XT}$ with
\[	\hat{p}^*_{XT}=\frac{\theta_X|\mathcal{I}^*_{XT}|}{\sum_X\theta_X|\mathcal{I}^*_{XT}|}	\]
where $\theta_X=|\mathcal{I}_{XT}|/|\mathcal{I}^*_{XT}|$ adjusts the weight given to each $X$ subgroup. The matched sample contains multiple pairs per original policy so the estimator is adjusted such that the effective weight given to each subgroup is the same as in the original portfolio. In the example in Section~\ref{sec:matchdemo}, if ties were included for policy l then all of the pairs (b,l), (c,l) and (g,l) would be included in the matched sample with a weight of $1/3$.

\section{Software implementation and cluster computation}\label{sec:softimp}
This section decribes the R code for implementing the matching methods, including the parallel computations and propensity score estimation. The \textit{Matching} and \textit{parallel} packages are required and need to be loaded before running the code. 

This section loads the dataset, which is a comma delimited file extracted from the \textit{CASdatasets} package. It also converts the vehicle power from a factor to a numeric variable so that it can be approximately matched.

\begin{verbatim}
premdata <- read.csv("fremotor1prem.csv")                 # csv dataset
premdata$tr <- ifelse(premdata$Year==2003,FALSE,TRUE)     # relabel year
set.seed(500)
premdata <- premdata[sample(1:nrow(premdata),size=5e3),]  # random data subset

# convert Vehicle Power to numeric
premdata$VehPower <- as.numeric(gsub("P","",
                        levels(premdata$VehPower)))[premdata$VehPower] 
\end{verbatim}

This code estimates the propensity score with a logistic regression and extracts the linear predictor to match.

\begin{verbatim}
pmodel <- glm(tr~DrivAge+DrivGender+BonusMalus+VehAge+VehPower # logistic
                  +VehClass+VehNb+Region+Garage,               # regression
                family=binomial,data=premdata)                 #     

pscore.lin <- pmodel$linear.predictor	                         # add linear 
premdata <- cbind(premdata,pscore.lin)                         # predictor
rm(pmodel,pscore.lin)                                          # to dataset
\end{verbatim}

Multiple matching methods are implemented below, with parallel computation used to reduce the computing time for the computational matching.

\begin{verbatim}
attach(premdata)		
X <- cbind(pscore.lin,DrivAge,DrivGender,BonusMalus,VehAge,VehPower, # variables
            VehClass,VehNb,Region,Garage)                            # to match

# classic matching on all X
     mtchout <- Match(Tr=tr,X=X,exact=c(rep(FALSE,6),rep(TRUE,4)),   
                          ties=FALSE,replace=FALSE)               

# complete matching on all X     
mtchout.exct <- Match(Tr=tr,X=X,exact=rep(TRUE,10),                  
                          ties=FALSE,replace=FALSE)

# match on PScore, exactly on critical variables
 mtchout.psc <- Match(Tr=tr,X=cbind(pscore.lin,VehClass,VehNb,Region,Garage),
                        exact=c(rep(FALSE,1),rep(TRUE,4)),                   
                          ties=FALSE,replace=FALSE)                          
 
cl <- makeCluster(detectCores()); cl # make/activate SOCK cluster

# computational matching, start at PScore matching
genout <- GenMatch(Tr=tr,X=X,BalanceMatrix=X,
                    exact=c(rep(FALSE,6),rep(TRUE,4)),ties=FALSE,replace=FALSE,
                    pop.size=2000,max.generations=500,wait.generations=5,
                    hard.generation.limit=TRUE,cluster=cl,
                    starting.values=c(1000,rep(1,5),rep(1000,4)))
stopCluster(cl) # stop cluster

mtchout.gen <- Match(Tr=tr,X=X,exact=c(rep(FALSE,6),rep(TRUE,4)),
                         ties=FALSE,replace=FALSE,Weight.matrix=genout)
detach(premdata)
\end{verbatim}

The optimal weights chosen by the genetic algorithm are (49, 442, 3, 1, 922, 655, 513, 571, 518, 775).


\begin{thebibliography}{}

\bibitem[{Abadie and Imbens(2006)}]{AI:2006}
Abadie, A. and Imbens, G.W. (2006).
\newblock {Large sample properties of matching estimators for average treatment effects}.
\newblock \textit{Econometrica}. \textbf{74} 1, 235--267.

\bibitem[{Abadie and Imbens(2016)}]{AI:2016}
Abadie, A. and Imbens, G.W. (2016).
\newblock {Matching on the estimated propensity score}.
\newblock \textit{Econometrica}. \textbf{84} 2, 781--807.

\bibitem[{Austin(2009)}]{A:2009}
Austin, P.C. (2009).
\newblock {Balance diagnostics for comparing the distribution of baseline covariates between treatment groups in propensity score matched samples}.
\newblock \textit{Statistics in Medicine}. \textbf{28}, 3083--3107.

\bibitem[{Austin and Small(2014)}]{AS:2014}
Austin, P.C. and Small, D.S. (2014).
\newblock {The use of botstrapping when using propensity score matching without replacement: a simulation study}.
\newblock \textit{Statistics in Medicine}. \textbf{33}, 4306--4319.

\bibitem[{Bodoff(2009)}]{B:2009}
Bodoff, N.M. (2009).
\newblock {Measuring rate change}.
\newblock \textit{Casualty Actuarial Society E-forum}. \textbf{Winter} 2009.

\bibitem[{Brookhart et al.(2006)}]{BSRGAS:2006}
Brookhart, M.A., Schneeweiss, S., Rothman, K.J., Glynn, R.J., Avorn, J. and St\"urmer, T. (2006).
\newblock {Variable selection for propensity score models}.
\newblock \textit{American Journal of Epidemiology}. \textbf{163}, 1149-1156.

\bibitem[{Dawid(2002)}]{D:2002}
Dawid, A.P. (2002).
\newblock {Influence diagrams for causal modelling and inference}.
\newblock \textit{International Statistical Review}. \textbf{70} 2, 161--189.

\bibitem[{Dutang and Charpentier(2018)}]{DC:2018}
Dutang, C. and Charpentier, A. (2018).
\newblock {CASdatasets: insurance datasets}.
\newblock \textit{R package 1.0-8}. \texttt{http://dutangc.free.fr/pub/RRepos/web/CASdatasets-index.html}.

\bibitem[{England and Verrall(2002)}]{EV:2002}
England, P.D. and Verrall, R.J. (2002).
\newblock {Stochastic claims reserving in general insurance}.
\newblock \textit{British Actuarial Journal}. \textbf{8} 3, 443--518.

\bibitem[{Greenland et~al.(1999)}]{GRP:1999}
Greenland, S., Robins, J.M. and Pearl, J. (1999).
\newblock {Confounding and collapsibility in causal inference}.
\newblock \textit{Statistical Science}. \textbf{14} 1, 29--46.

\bibitem[{Lloyd's(2018)}]{L:2018}
Lloyd's (2018).
\newblock {Performance management data return instructions 2018 v1.1}.
\newblock \texttt{https://www.\\lloyds.com/market-resources/underwriting/performance-management-data-return-\\pdmr}.

\bibitem[{Hansen(2008)}]{H:2008}
Hansen, B.B. (2008).
\newblock {The prognostic analogue of the propensity score}.
\newblock \textit{Biometrika}. \textbf{95} 2, 481--488.

\bibitem[{Heckman et~al.(1998)}]{HIT:1998}
Heckman, J.J., Ichimura, H. and Todd, P. (1998).
\newblock {Matching as an econometric evaluation estimator}.
\newblock \textit{Review of Economic Studies}. \textbf{65}, 261--294.

\bibitem[{Ho et~al.(2007)}]{HIKS:2007}
Ho, D.E., Imai, K., King, G. and Stuart, E.A. (2007).
\newblock {Matching as nonparametric preprocessing for reducing model dependence in parametric causal inference}.
\newblock \textit{Political Analysis}. \textbf{15}, 199--236.

\bibitem[{McClenahan(2001)}]{M:2001}
McClenahan, C. (2001).
\newblock {Ratemaking}.
\newblock \textit{Foundations of Casualty Actuarial Science}. \textbf{4th} edition.

\bibitem[{R Core Team(2017)}]{RCT:2017}
R Core Team (2017).
\newblock {R: a language and environment for statistical computing}.
\newblock \textit{R Foundation for Statistical Computing}, Vienna, Austria. \texttt{https://www.R-project.org/}.

\bibitem[{Ramsahai et~al.(2011)}]{RGS:2011}
Ramsahai, R.R., Grieve, R.G. and Sekhon, J.S. (2011).
\newblock {Extending iterative matching methods: an approach to improving covariate balance that allows prioritisation}.
\newblock \textit{Health Services and Outcomes Research Methodology}. \textbf{11}, 95--114.

\bibitem[{Robins et~al.(2000)}]{RHB:2000}
Robins, J.M., Hernan, M.A. and Brumback, B. (2000).
\newblock {Marginal structural models and causal inference in epidemiology}.
\newblock \textit{Epidemiology}. \textbf{11}, 550--560.

\bibitem[{Rosenbaum and Rubin(1983)}]{RR:1983}
Rosenbaum, P.R. and Rubin, D.B. (1983).
\newblock {The central role of the propensity score in observational studies for causal effects}.
\newblock \textit{Biometrika}. \textbf{70}, 410--455.

\bibitem[{Rosenbaum and Rubin(1985)}]{RR:1985}
Rosenbaum, P.R. and Rubin, D.B. (1985).
\newblock {Constructing a control group using multivariate matched sampling methods that incorporate the propensity score}.
\newblock \textit{The American Statistician}. \textbf{39} 1, 33--38.

\bibitem[{Rubin(1980)}]{R:1980}
Rubin, D.B. (1980).
\newblock {Bias reduction using Mahalanobis-metric matching}.
\newblock \textit{Biometrics}. \textbf{36}, 293--298.

\bibitem[{Sekhon(2011)}]{S:2011}
Sekhon, J.S. (2011).
\newblock {Multivariate and propensity score matching software with automated balance optimization: the matching package for R}.
\newblock \textit{Journal of Statistical Software}. \textbf{42}, 1--52.

\bibitem[{Stuart(2010)}]{S:2010}
Stuart, E.A. (2010).
\newblock {Matching methods for causal inference: a review and a look forward}.
\newblock \textit{Statistical Science}. \textbf{25} 1, 1--21.

\bibitem[{Vaughn(2004)}]{V:2004}
Vaughn, T. (2004).
\newblock {Commercial lines price monitoring}.
\newblock \textit{CAS Forum}. 497--519.


\end{thebibliography}
\end{document}